\documentclass[12pt]{article}

\usepackage{graphicx}
\usepackage{subcaption}
\usepackage{times}
\usepackage[backend=biber,
            style=numeric,
            sorting=none]{biblatex}
\newenvironment{sciabstract}{%
\begin{quote} \bf}
{\end{quote}}

\title{Photonic classification on a single diffractive layer} 

\author
{Anil J. Pekgöz,$^{1,2,\ast}$ Emre Yüce,$^{1,\ast}$ \\
\\
\normalsize{$^{1}$ Programmable Photonics Group, Department of Physics, Middle East Technical University,}\\
\normalsize{06800 Ankara, Turkey}\\
\normalsize{$^{2}$ Quantum Science and Technology, School of Natural Science, }\\
\normalsize{Technische Universität München, 85748 Garching, Germany}\\
\\
\normalsize{$^\ast$Corresponding Author: anil.pekgoez@tum.de}\\
\normalsize{\hspace{121pt} eyuce@metu.edu.tr}
}

\date{}

\begin{document} 


\baselineskip24pt


\maketitle


\begin{sciabstract}
  Photonic computation started to shape the future of fast, efficient and accessible computation. The advantages brought by light based Diffractive Deep Neural Networks (D2NN), are shown to be overwhelmingly advantageous especially in targeting classification problems. However, cost and complexity of multi-layer systems are the main challenges that reduce the deployment of this technology. In this study, we develop a simple yet extremely efficient way to achieve optical classification using a single diffractive optical layer. A spatial light modulator is used not only to emulate the classifying system but also the input medium. We show that using a simple interpretable linear classifier, images can be classified at the speed of light. We perform classification of road signs  under the effect of noise and demonstrate that we can successfully classify input images with more than 90$\%$ accuracy even with 13$\%$ noise/imperfection. 
\end{sciabstract}
\newpage

\section*{Introduction} 
The penetration of Artificial Intelligence (AI) to our daily routine tasks increases at an unprecedented rate. Heavily demanded cognitive copilots that are mainly based on AI need further improvement especailly in energy efficiency. The efficiency aspect of the new AI systems is important for the sustainability purposes besides the efforts spent for improving the capabilities of new generation AI systems. 
\vspace{12pt}

The imposed low parallelizability \cite{vegh-2019} (layer by layer computation) of Deep Neural Networks (DNN) using conventional processors is a bottle neck for AI \cite{desislavov-2023}. In many applications, the need for immediate and energy efficient response is increasing. For this reason, the developers are seeking for alternative basis for Neural Networks. One of the alternatives is the Diffractive Deep Neural Network (D$^2$NN) which is based on the light diffraction theory \cite{goodman-2005} and efficiently emulates a Neural Networks. The advantages of D$^2$NN systems are: power efficiency \cite{caulfield-1989} and the incredible speeds \cite{yu-1993} relative to the common Deep Neural Networks which are based on silicon processors. 
\vspace{12pt}

Pioneering studies have shown that it is possible to design a system \cite{zuo-2021} that is based only on Spatial Light Modulators (SLM) to simulate neurons and their interconnections to classify Modified National Institute of Standards and Technology database (MNIST database) numbers \cite{deng-2012} and fashion objects \cite{xiao-2017}. Moreover, researchers \cite{zhou-2021} also have shown that numbers can be classified with only a Digital Micromirror Device, as the input layer, and a SLM, as the classifying diffractive layer. The potential of easy applicability of D$^2$NN systems have been proven \cite{lin-2018} in a system of 3D printed Diffractive Optical Elements (DOE) where each DOE layer acts as a neural layer. 
\vspace{12pt}

However, the multi-layer D$^2$NN systems \cite{zuo-2021, zhou-2021,lin-2018, liu-2022} inherintly employ layer-wise complexity in their structure. This complexity limits the fields of application where shorter optimization time and cost-effectiveness is required. Moreover, the correlation of training time and overall network complexity limits scalability \cite{paola-1997}. As the number of layers increase the interpretability of the solution decreases and the data becomes linearly inseparable. The risk of overfitting also increases by increasing number of layers and can be extremely inefficient with a small number of dataset. 
\vspace{12pt}

With this challenges in mind, we develop a true linear photonic optical classifier where the input and classifier layer are combined in a single diffractive layer on a SLM surface. Since the input and classifier layer are combined on the same plane, the theoretical and methodological approaches are simplified substantially and prove optical classification without the need for deep layers. Since our classification method does not require any feature transformation the output is deterministic, interpretable and linearly separable. Our optimization algorithm is mostly based on the Gerchberg-Saxton algorithm \cite{gerchberg-1972, poon-2014} with a basic spatial restriction. The performance of our system is demonstrated by diagonality of test, prediction histograms and confusion matrices. The performance of our method is not deteriorated in the presence of noise unlike the Neural Networks that are intolerant to noise \cite{paranhos-2016,dodge-2016}. 
\vspace{12pt}


\section*{Results and Discussion}
\begin{figure}[h]
\centering
\begin{subfigure}{.55\textwidth}
  \centering
  \includegraphics[width=1\linewidth]{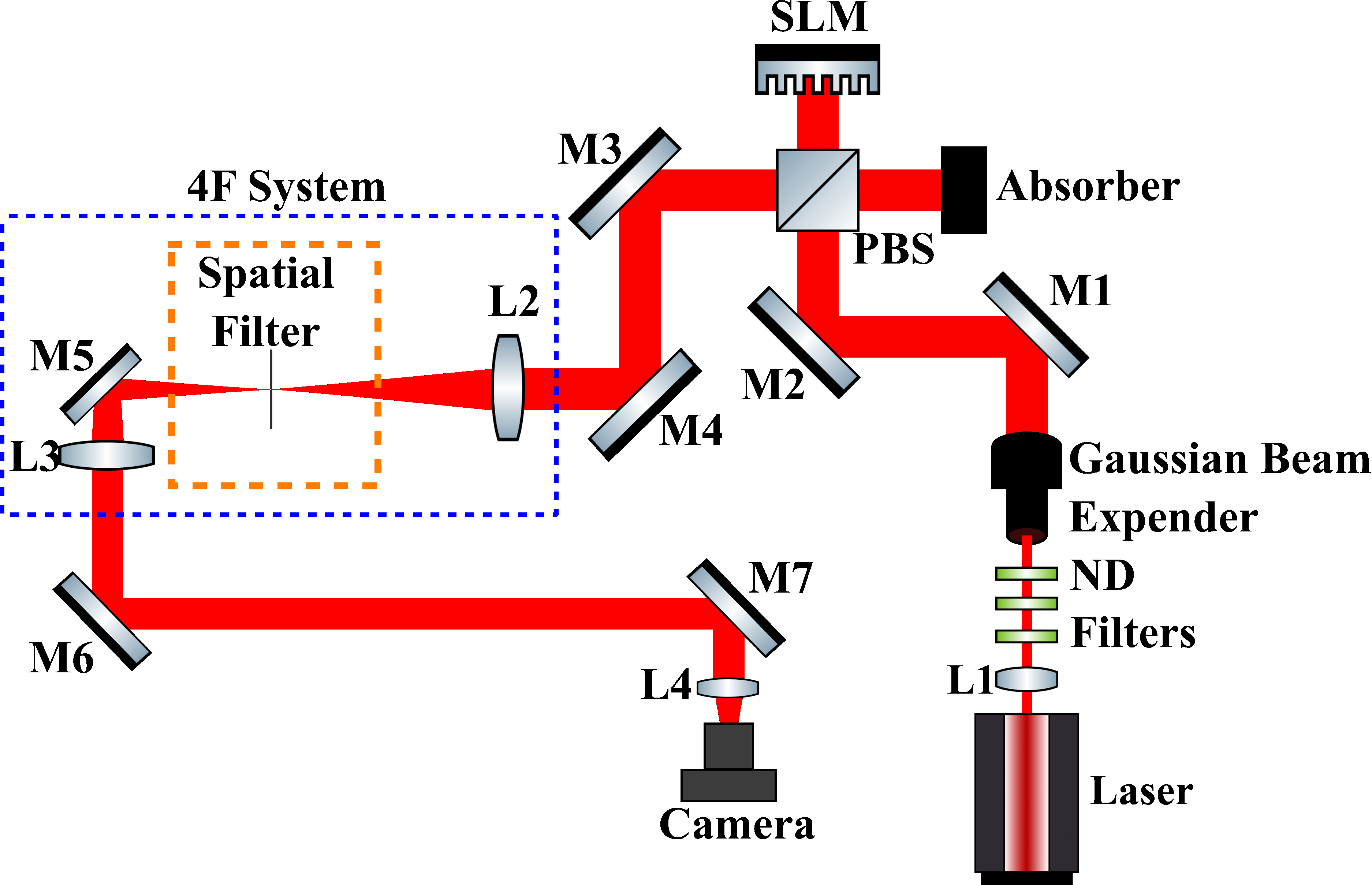}
\end{subfigure}%
\begin{subfigure}{.5\textwidth}
  \centering
  \includegraphics[width=.60\linewidth]{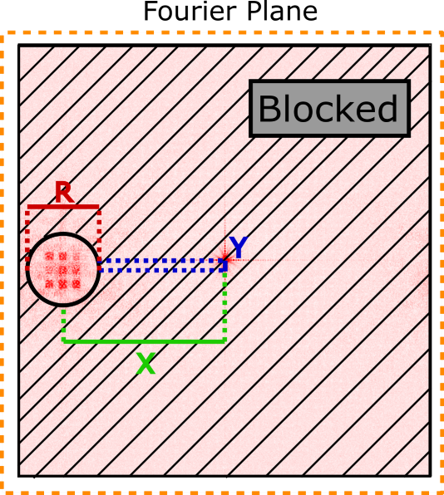}
\end{subfigure}
\caption{\textbf{A,} The schematic illustration of the optical system. \textbf{B,} The Fourier Plane at the Spatial Filter. Only a circular region at the position (X,Y) with a diameter of R is allowed through the Spatial Filter. L: Lens, M: Mirror, ND: Neutral Density, PBS: Polarizing Beam Splitter, SLM: Spatial Light Modulator.}
\label{Fig1}
\end{figure}

Our linear classifier is based on the system given in  Fig. \ref{Fig1}A. The light source is a semiconductor diode laser at $\lambda=632~\rm{nm}$ and is attenuated with a set of ND filters. The laser beam is expanded using a Gaussian Beam Expander in order to increase the coverage at the surface of the SLM. The Polarizing Beam Spliter (PBS) acts both as a redirector for the reflected beam from the SLM and as a polarizer which aligns the polarization of the laser beam to the SLM's modulation axis. The modulated beam is then redirected to the 4f system where only a small region of the frequency space is let through the iris, see Fig. \ref{Fig1}B. The filtered laser beam is then redirected with multiple mirrors into the last lens in front of the camera which performs a Fourier transformation of the input field\cite{goodman-2005} .

\begin{figure}[h]
    \centering
    \includegraphics[width=0.8\linewidth]{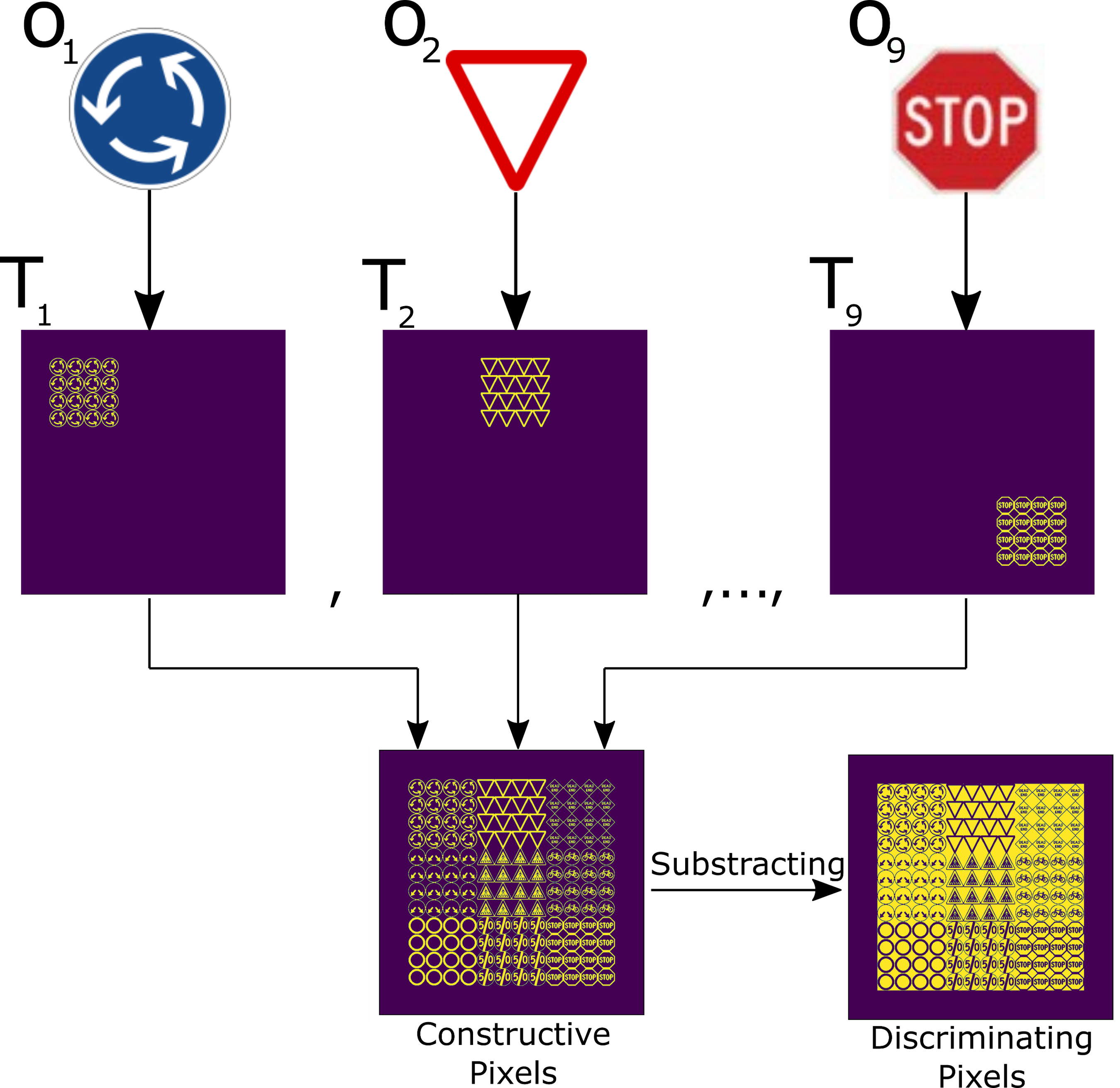}
    \caption{The grid-wise multiples of the object's binary image is placed into the corresponding grouped-surface (9 grouped-surfaces). Each grouped-surface consists of 16 replicas of the input image. The pixels that are outside the input images are called as "Discriminating pixels" (illustrated by subtracting the input images from the phase mask) and are shown in bottom right corner.}
    \label{Fig2}
\end{figure}

The surface of the SLM is spatially divided into grouped-surfaces according to the number of objects. This approach is used to optimize each object individually to it's respective grouped-surface. Since the grouped-surfaces are spatially divided, they do not interact with each other on the surface level at the SLM. The binary shapes of the objects are written onto those grouped-surfaces to establish the spatial restriction on the Gerchberg-Saxton Algorithm. To increase the used space on each grouped-surface, the binary image of the object is written in a multiple-grid shape manner. The combined area which is going to be used in optimization process is named as "Classifying Pixels". The area which is outside Constructive Pixels is used to diverge the light power outside the target and is named as "Discriminating Pixels". This process of dividing the surface of the SLM and writing the binary object as grid-wise manner is shown in Fig. \ref{Fig2}. Given the circular nature of the Laser Beam, the center $600 \times 600$ pixels of the SLM are effectively used in our classification process.

The spatial restriction indicated in Fig. \ref{Fig2}. is used as a constraint at the image plane for the Gerchberg-Saxton Algorithm in each iteration \cite{poon-2014}. The phase optimization is only performed on the spatially restricted regions which are used for phase optimization on the SLM. The area is filled with the  binarized form  of the 9 standard traffic signs \cite{chaki-2014}. The mathematical representation of the process is as follows:

$$\Psi_{0, i}(x,y) = A_0(x,y)\exp{(i\phi'_{0,i}(x,y))}$$,

where $\Psi_{0,i}$ is the subwavefront of $0^{th}$ iteration of the $i^{th}$ object from the restricted SLM space and $A(x,y)$ is the amplitude distribution of the laser's wavefront. The $\phi'_{0,i}(x,y)$ is the restricted phase mask, which is defined as:

$$\phi'_{0, i}(x,y) = B_i(x,y)\odot\phi_{0,i}(x,y)$$.

\begin{figure}[h]
    \centering
    \includegraphics[width=0.95\linewidth]{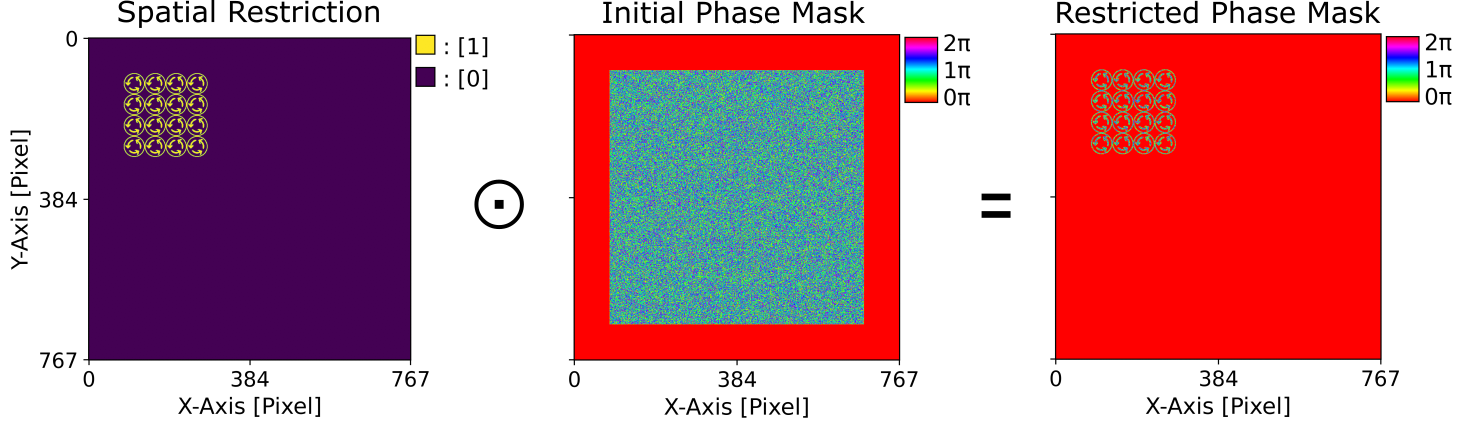}
    \caption{The binary form of the Spatial Restriction Array is used in element-wise multiplication to include only the "Constructive Pixels" of the object which is optimized via the Gerchberg-Saxton Algorithm. The restricted mask that is used for iterative optimization is shown in the right panel. The "Discriminating pixels Fig. \ref{Fig2} on the other hand are used for improving the signal to noise level that increases the classification success rate. In this way, a larger pixel density is engaged for the classification.}
    \label{Fig3}
\end{figure}

\noindent The $B_i(x,y)$ is the binary form of the Spatial Restriction Array and $\phi_{0,i}(x,y)$ is the initial phase mask, which is generated randomly at the beginning of the process. This part is depicted visually in Fig. \ref{Fig3}. The $B_i(x,y)$ is divided in a $4 \times 4$ grid with the $i^{th}$ object. With 9 objects, the SLM is divided in $12 \times 12$ grids in total. This results in a $200 \times 200$ pixels of Spatial Restriction Array with $50 \times 50$ pixels of single image resolution. 


The optimization of each grouped-surface's phase mask, $\phi'_{0, i}(x,y)$, is performed with 20 iteration of the Spatially Restrictive Gerchberg-Saxton algorithm. This results in a total of 200 iteration including the "Discriminating Pixels" surface. The total duration of one time optimization effort took 20 minutes with a single core (Intel i7-10750H) processor. Given that the grouped-surfaces are independent from each other, even the one time optimization duration could be shortened to almost 4.8 minutes with 6 parallel cores and a paralellity fraction of $p \approx 0.95$ according to Amdahl's Law \cite{reddy-2011, bryant-2019}, see Methods Section for details of the algorithm. In addition, the main time limitation arise from the frame rate of camera and mostly by the refresh rate of the SLM (60 Hz). 

\begin{figure}[h]
\centering
\begin{subfigure}{.5\textwidth}
  \centering
  \includegraphics[width=1.0\linewidth]{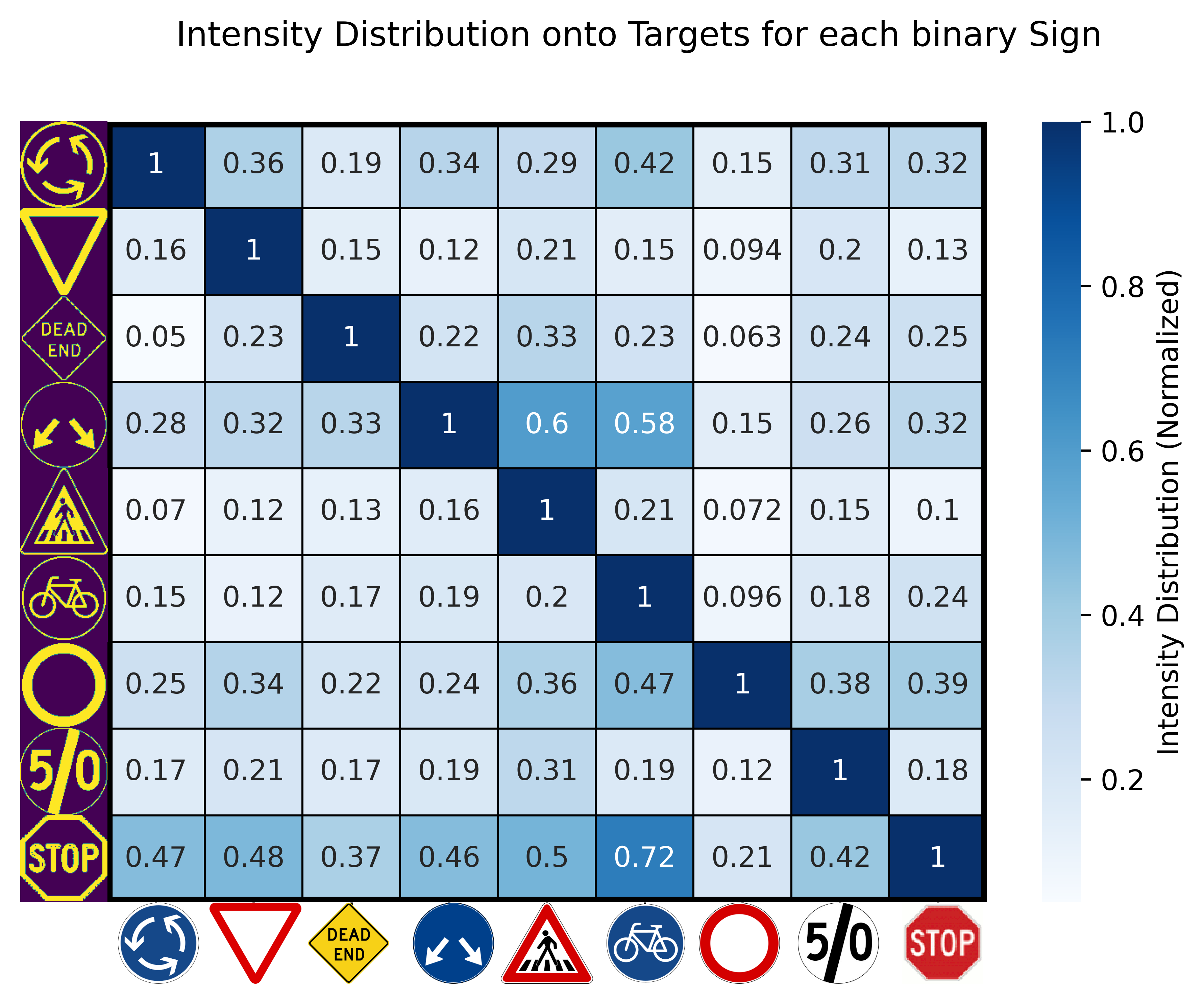}
\end{subfigure}%
\begin{subfigure}{.5\textwidth}
  \centering
  \includegraphics[width=1\linewidth]{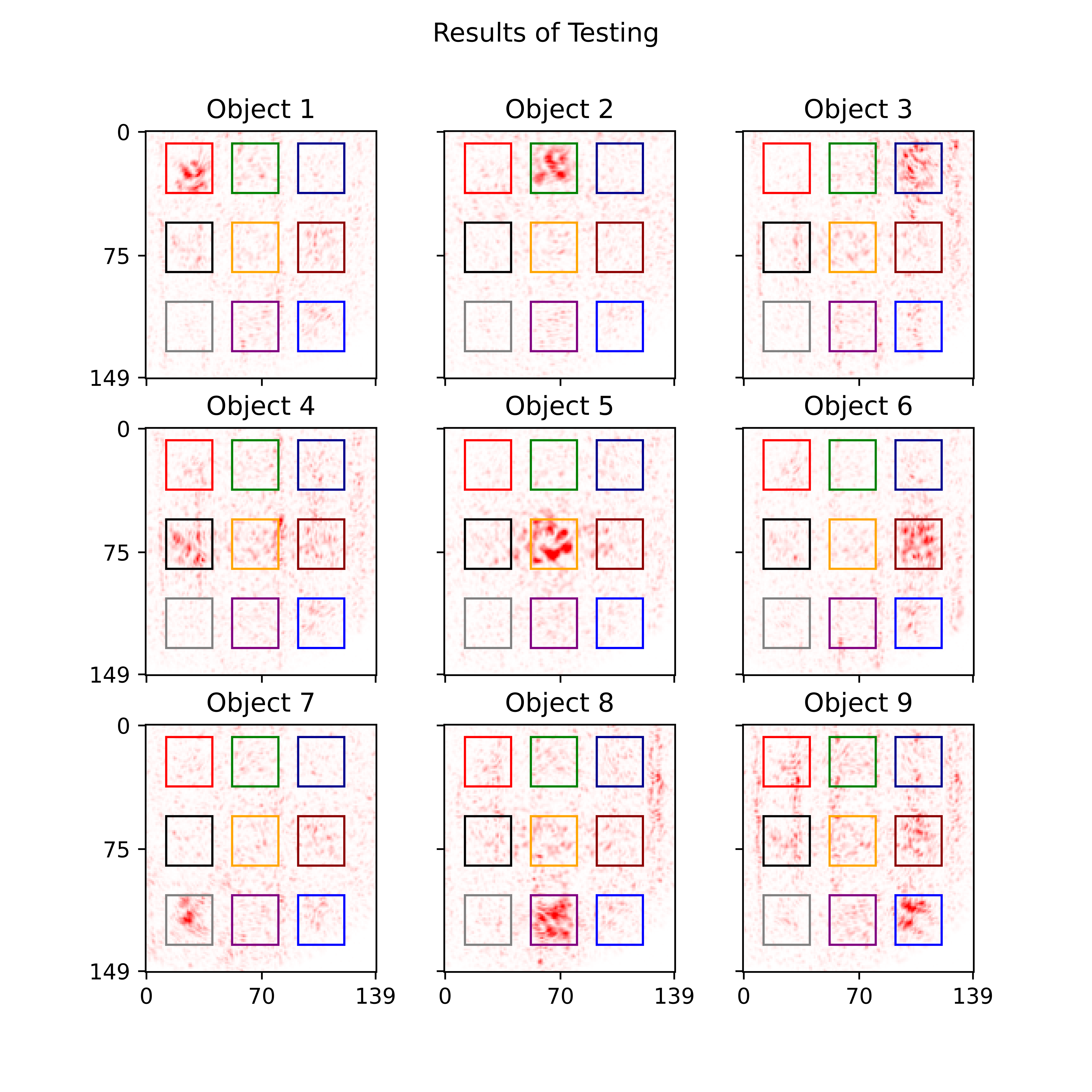}
\end{subfigure}
\caption{\textbf{A,} The confusion matrix indicating the performance of the classification. The intensity distribution is shown as a matrix form where each row corresponds to the histographic intensity distribution and the values are normalized to the maximum value. \textbf{B,} The targets used in the optimization process are $3\times3$ grid with constant height, width, horizontal and vertical separation. The coordinate system of the Fourier Plane in the algorithm and the coordinate system of the real Fourier Plane on the camera are not mapped exactly due to the magnification of the 4f system and possible slight misplacement of the lenses. Therefore, there is a slight linear transformation between those Fourier Planes. }
\label{Fig4}
\end{figure}

The power distribution on the camera is used as a performance indicator for the resulting phase plates. Every initial object is grided as $12 \times 12$ to cover the surface of our SLM. The normalized histogram results are shown in Fig. \ref{Fig4}A. The diagonallity indicates the successful operation of the phase plate that focuses the laser beam onto the designated region that are shown in Fig. \ref{Fig4}B. From the histogram in Fig. \ref{Fig4}A and the images of the camera shown in Fig. \ref{Fig4}B, it is clearly visible that the laser is majorly concentrated on the designated target with some small signal outside the target area. In our experimental setup, we used the Holoeye LC-R 720 as our SLM. The phase shift capability in the specific configuration is 1.81$\pi$. This fact causes an inefficient wavefront shaping since the whole phase range of 2$\pi$ was not available \cite[p.~186]{poon-2014}. Therefore, with a better phase shift range, higher validation percentages could be reached. Moreover, the repetition of the images in a grid-wise manner acts as a grating which produces unintentionally an interference pattern which overlaps the target areas. This interference pattern could be removed via modification of the "Discriminating Pixels" such that it interferes destructively with the patterns in question.

The one-to-one linear scaling of the Image Plane and the Fourier Plane causes a significant problem on low resolution work space on a SLM. This problem is the precision and resolution of the generated target areas since it is bounded on the Image Plane's size and resolution. A smaller target size increases the validation efficiency and could also be scaled up with an optical magnifier system. However, the limit of the target size is bounded on the lowest achievable pixel size in the optimization algorithm which is also bounded by the SLM. To test the validation performance at different noise levels, 20 datasets have been generated with a predefined percentage of random noise. Each dataset has 9000 images which are equally distributed among 9 objects. The noise percentage ranges from 1$\%$ to 20$\%$ with a step size of 1$\%$. Three resulting confusion matrices for 1$\%$, 5$\%$ and 10$\%$ and corresponding samples are shown in Fig. \ref{Fig5}A. With this algorithm, we managed to reach to a validation percentage of 96.26$\%$ for 10$\%$ noise and a validation rate of 74.66$\%$ for 20$\%$ noise. We can archive a validation rate more then 90$\%$ till 13$\%$ noise level. The combined validation average of 20 datasets consisting of 180000 images with various noise levels equals to an average of 91.78$\%$.

The single layer nature of our proposed system showed us the possibility of optical classification via a novel method which is not directly connected to the fundamental working principles of interconnected neurons which spans over multiple layers like in DNN systems \cite{schmidhuber-2014} and a D$^2$NN systems. It is important to state that this new approach could be also adopted to the silicon based computation. 

Our method provides simple yet novel way to classify objects. The nature of not using any deep layers greatly simplifies system structure. Moreover, the system can be further improved with an end classifier to get better noise vs validation performance results. Since our method is simple and linearly separable it can form a basis for more efficient classifying algorithms that require efficient and fast response times. We think that affordable single layer classification of traffic signs will have a major safety role for the increasing number of autonomous vehicles that are being deployed to our daily routine. The affordability and simplicity of the method will encourage manufacturers to deploy safer products into the market. 


\section*{Methods and Materials}
\textbf{Spatially Restrictive Gerchberg-Saxton Algorithm}

In the classification process, a Gerchberg-Saxton Algorithm is used to optimize the surface of the SLM. The property of declaring a spatial, shape-wise restriction on the Gerchberg-Saxton Algorithm is used as the primary optimization factor that enables us to classify various objects based on the shape-based characteristics of the object.

\noindent \textbf{Noise Addition Process and Testing}

To test the performance of our method, a synthetic, randomly-noise-induced datasets with various level of noise have been created. The noise adding process, depicted as in Suppl. Fig. \ref{suplFig1}A, flips the values of the binary image either constructively [0 → 1] or destructively [0 → 1]. The position of the noisy pixels were selected randomly and selected once. With this rule, double selection (i.e. 0 → 1 → 0) is prevented. To test each noise induced image, it is grid wise multiplied to cover the whole sub-surfaces available on the SLM. This grid wise multiplication is depicted on Suppl. Fig. \ref{suplFig1}B. The final binary mask is then element wise multiplied with the final trained phase mask and tested on the SLM. 


\printbibliography

@article{vegh-2019,
	author = {given-i=J, given=János, family=Végh},
	date = {2019-04-11},
	doi = {10.1186/s40708-019-0097-2},
	journaltitle = {Brain Informatics},
	number = {1},
	title = {How Amdahl’s Law limits the performance of large artificial neural networks},
	url = {https://doi.org/10.1186/s40708-019-0097-2},
	volume = {6},
}

@article{desislavov-2023,
	author = {given-i=R, given=Radosvet, family=Desislavov and given-i=F, given=Fernando, family=Martínez-Plumed and given-i=J, given=José, family=Hernández-Orallo},
	date = {2023-02-26},
	doi = {10.1016/j.suscom.2023.100857},
	journaltitle = {Sustainable Computing Informatics and Systems},
	pages = {100857},
	title = {Trends in AI inference energy consumption: Beyond the performance-vs-parameter laws of deep learning},
	url = {https://doi.org/10.1016/j.suscom.2023.100857},
	volume = {38},
}

@book{goodman-2005,
	author = {given-i=JW, given={Joseph W.}, family=Goodman},
	date = {2005-01-01},
	publisher = {Roberts and Company Publishers},
	title = {Introduction to Fourier Optics},
}

@article{caulfield-1989,
	author = {given-i=H, given=H.J., family=Caulfield and given-i=J, given=J., family=Kinser and given-i=S, given=S.K., family=Rogers},
	date = {1989-01-01},
	doi = {10.1109/5.40669},
	journaltitle = {Proceedings of the IEEE},
	number = {10},
	pages = {1573--1583},
	title = {Optical neural networks},
	url = {https://doi.org/10.1109/5.40669},
	volume = {77},
}

@book{yu-1993,
	author = {given-i=FT, given={Francis T.S.}, family=Yu},
	booktitle = {Progress in optics},
	date = {1993-01-01},
	doi = {10.1016/s0079-6638(08)70162-8},
	pages = {61--144},
	title = {II Optical Neural Networks: architecture, design and models},
	url = {https://doi.org/10.1016/s0079-6638(08)70162-8},
}

@article{zuo-2021,
	author = {given-i=Y, given=Ying, family=Zuo and given-i=Y, given=Yujun, family=Zhao and given-i=Y, given=You-Chiuan, family=Chen and given-i=S, given=Shengwang, family=Du and given-i=J, given=Junwei, family=Liu},
	date = {2021-05-17},
	doi = {10.1103/physrevapplied.15.054034},
	journaltitle = {Physical Review Applied},
	number = {5},
	title = {Scalability of All-Optical neural networks based on spatial light modulators},
	url = {https://doi.org/10.1103/physrevapplied.15.054034},
	volume = {15},
}

@article{deng-2012,
	author = {given-i=NL, given={Li}, family=Deng},
	date = {2012-10-19},
	doi = {10.1109/msp.2012.2211477},
	journaltitle = {IEEE Signal Processing Magazine},
	number = {6},
	pages = {141--142},
	title = {The MNIST Database of Handwritten Digit Images for Machine Learning Research [Best of the Web]},
	url = {https://doi.org/10.1109/msp.2012.2211477},
	volume = {29},
}

@article{xiao-2017,
	author = {given-i=H, given=Han, family=Xiao and given-i=K, given=Kashif, family=Rasul and given-i=R, given=Roland, family=Vollgraf},
	date = {2017-01-01},
	doi = {10.48550/arxiv.1708.07747},
	journaltitle = {arXiv (Cornell University)},
	title = {Fashion-MNIST: a Novel Image Dataset for Benchmarking Machine Learning Algorithms},
	url = {https://arxiv.org/abs/1708.07747},
}

@article{zhou-2021,
	author = {given-i=T, given=Tiankuang, family=Zhou and given-i=X, given=Xing, family=Lin and given-i=J, given=Jiamin, family=Wu and given-i=Y, given=Yitong, family=Chen and given-i=H, given=Hao, family=Xie and given-i=Y, given=Yipeng, family=Li and given-i=J, given=Jingtao, family=Fan and given-i=H, given=Huaqiang, family=Wu and given-i=L, given=Lu, family=Fang and given-i=Q, given=Qionghai, family=Dai},
	date = {2021-04-12},
	doi = {10.1038/s41566-021-00796-w},
	journaltitle = {Nature Photonics},
	number = {5},
	pages = {367--373},
	title = {Large-scale neuromorphic optoelectronic computing with a reconfigurable diffractive processing unit},
	url = {https://doi.org/10.1038/s41566-021-00796-w},
	volume = {15},
}

@article{lin-2018,
	author = {given-i=X, given=Xing, family=Lin and given-i=Y, given=Yair, family=Rivenson and given-i=NT, given={Nezih T.}, family=Yardimci and given-i=M, given=Muhammed, family=Veli and given-i=Y, given=Yi, family=Luo and given-i=M, given=Mona, family=Jarrahi and given-i=A, given=Aydogan, family=Ozcan},
	date = {2018-07-26},
	doi = {10.1126/science.aat8084},
	journaltitle = {Science},
	number = {6406},
	pages = {1004--1008},
	title = {All-optical machine learning using diffractive deep neural networks},
	url = {https://doi.org/10.1126/science.aat8084},
	volume = {361},
}

@article{liu-2022,
	author = {given-i=C, given=Che, family=Liu and given-i=Q, given=Qian, family=Ma and given-i=ZJ, given={Zhang Jie}, family=Luo and given-i=QR, given={Qiao Ru}, family=Hong and given-i=Q, given=Qiang, family=Xiao and given-i=HC, given={Hao Chi}, family=Zhang and given-i=L, given=Long, family=Miao and given-i=WM, given={Wen Ming}, family=Yu and given-i=Q, given=Qiang, family=Cheng and given-i=L, given=Lianlin, family=Li and given-i=TJ, given={Tie Jun}, family=Cui},
	date = {2022-02-21},
	doi = {10.1038/s41928-022-00719-9},
	journaltitle = {Nature Electronics},
	number = {2},
	pages = {113--122},
	title = {A programmable diffractive deep neural network based on a digital-coding metasurface array},
	url = {https://doi.org/10.1038/s41928-022-00719-9},
	volume = {5},
}

@online{paola-1997,
	author = {given-i=JD, given={J. D.}, family=Paola and given-i=R, given=R., family=Schowengerdt},
	date = {1997},
	title = {The effect of Neural-Network structure on a multispectral Land-Use/Land-Cover classification},
	url = {https://api.semanticscholar.org/CorpusID:130948625},
}

@article{gerchberg-1972,
	author = {given-i=GR, given={Gerchberg R.}, family=W},
	date = {1972},
	journaltitle = {CiNii Research},
	title = {A practical algorithm for the determination of phase from image and diffraction plane pictures},
	url = {http://ci.nii.ac.jp/naid/10025518647/},
}

@book{poon-2014,
	author = {given-i=T, given=Ting-Chung, family=Poon and given-i=J, given=Jung-Ping, family=Liu},
	date = {2014-01-05},
	doi = {10.1017/cbo9781139061346},
	title = {Introduction to modern digital holography},
	url = {https://doi.org/10.1017/cbo9781139061346},
}

@online{paranhos-2016,
	author = {given-i=DCGB, given={Da Costa Gabriel B.}, family=Paranhos and given-i=WA, given={Welinton A.}, family=Contato and given-i=TS, given={Tiago S.}, family=Nazare and given-i=JESB, given={João E. S. Batista}, family=Neto and given-i=M, given=Moacir, family=Ponti},
	date = {2016-09-09},
	title = {An empirical study on the effects of different types of noise in image classification tasks},
	url = {http://arxiv.org/abs/1609.02781},
}

@inproceedings{dodge-2016,
	author = {given-i=S, given=Samuel, family=Dodge and given-i=L, given=Lina, family=Karam},
	date = {2016-06-01},
	doi = {10.1109/qomex.2016.7498955},
	pages = {1--6},
	title = {Understanding how image quality affects deep neural networks},
	url = {https://doi.org/10.1109/qomex.2016.7498955},
}

@book{chaki-2014,
	author = {given-i=N, given=Nabendu, family=Chaki and given-i=SH, given={Soharab Hossain}, family=Shaikh and given-i=K, given=Khalid, family=Saeed},
	booktitle = {Studies in computational intelligence},
	date = {2014-01-01},
	doi = {10.1007/978-81-322-1907-1_2},
	pages = {5--15},
	title = {A comprehensive survey on image binarization techniques},
	url = {https://doi.org/10.1007/978-81-322-1907-1_2},
}

@book{reddy-2011,
	author = {given-i=M, family=Reddy},
	date = {2011-01-01},
	doi = {10.1016/c2010-0-65832-9},
	title = {API design for C++},
	url = {https://doi.org/10.1016/c2010-0-65832-9},
}

@book{bryant-2019,
	author = {given-i=RE, given={Randal E.}, family=Bryant and given-i=DR, given={David R.}, family=O'Hallaron},
	date = {2019-07-12},
	publisher = {Pearson Higher Ed},
	title = {Computer Systems: A Programmer's Perspective, Global Edition},
}

@article{schmidhuber-2014,
	author = {given-i=J, given=Jürgen, family=Schmidhuber},
	date = {2014-10-13},
	doi = {10.1016/j.neunet.2014.09.003},
	journaltitle = {Neural Networks},
	pages = {85--117},
	title = {Deep learning in neural networks: An overview},
	url = {https://doi.org/10.1016/j.neunet.2014.09.003},
	volume = {61},
}
\clearpage

\newpage

\newpage

\newpage

\begin{figure}[h]
    \centering
   \includegraphics[width=1\linewidth]{Fig5.png}
    \caption{The noisy validation images are generated with the same algorithm shown in Figure 4 was used to generate a dataset of 9000 images (1000 for each object) for every noise percentages. The test results for 1$\%$, 5$\%$, 10$\%$ and 20$\%$ are shown with a sample from the noisy dataset and their corresponding Confusion Matrices. For 10$\%$ and 20$\%$ noise, the validation rate was 96.26$\%$ and 74.66$\%$ respectively.}
    \label{Fig5}
\end{figure}

\newpage

\renewcommand{\figurename}{Supplementary Figure}
\setcounter{figure}{0}

\begin{figure}[h]
\centering
\begin{subfigure}[b]{0.7\textwidth}
   \includegraphics[width=1\linewidth]{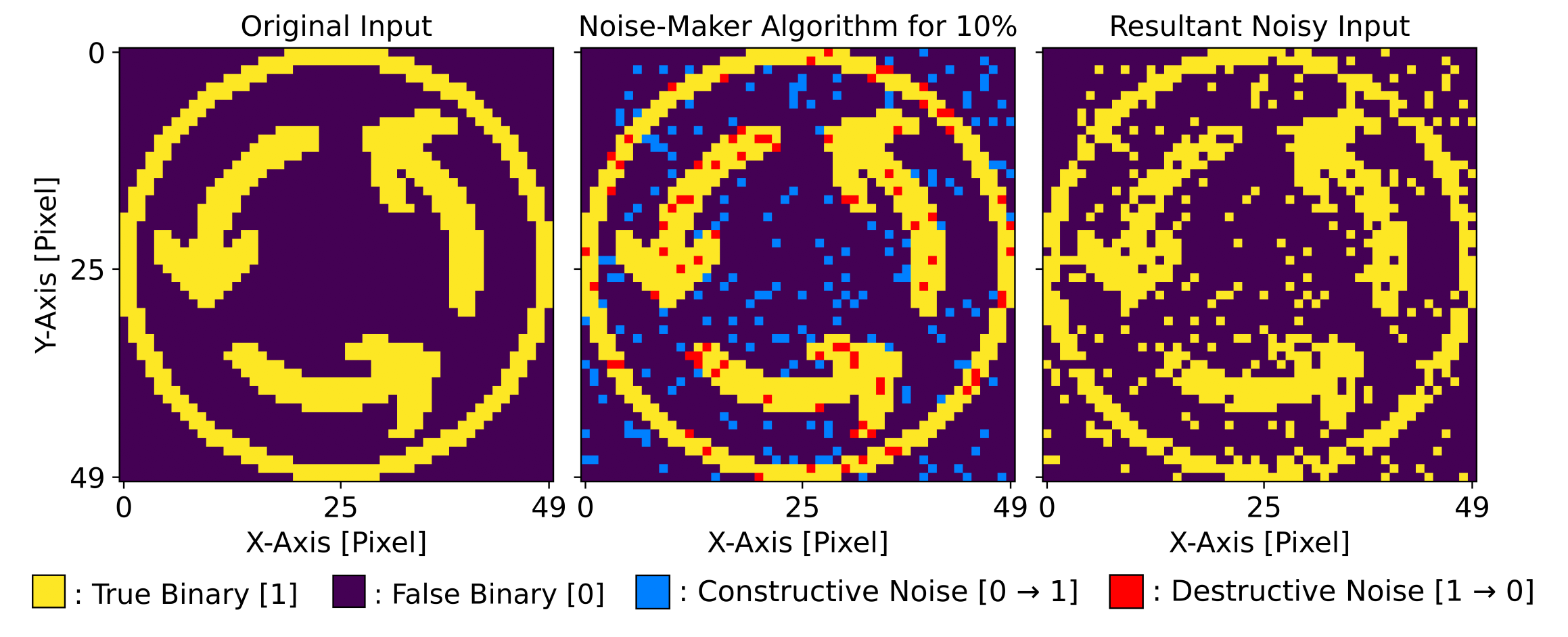}
\end{subfigure}

\begin{subfigure}[b]{0.7\textwidth}
   \includegraphics[width=1\linewidth]{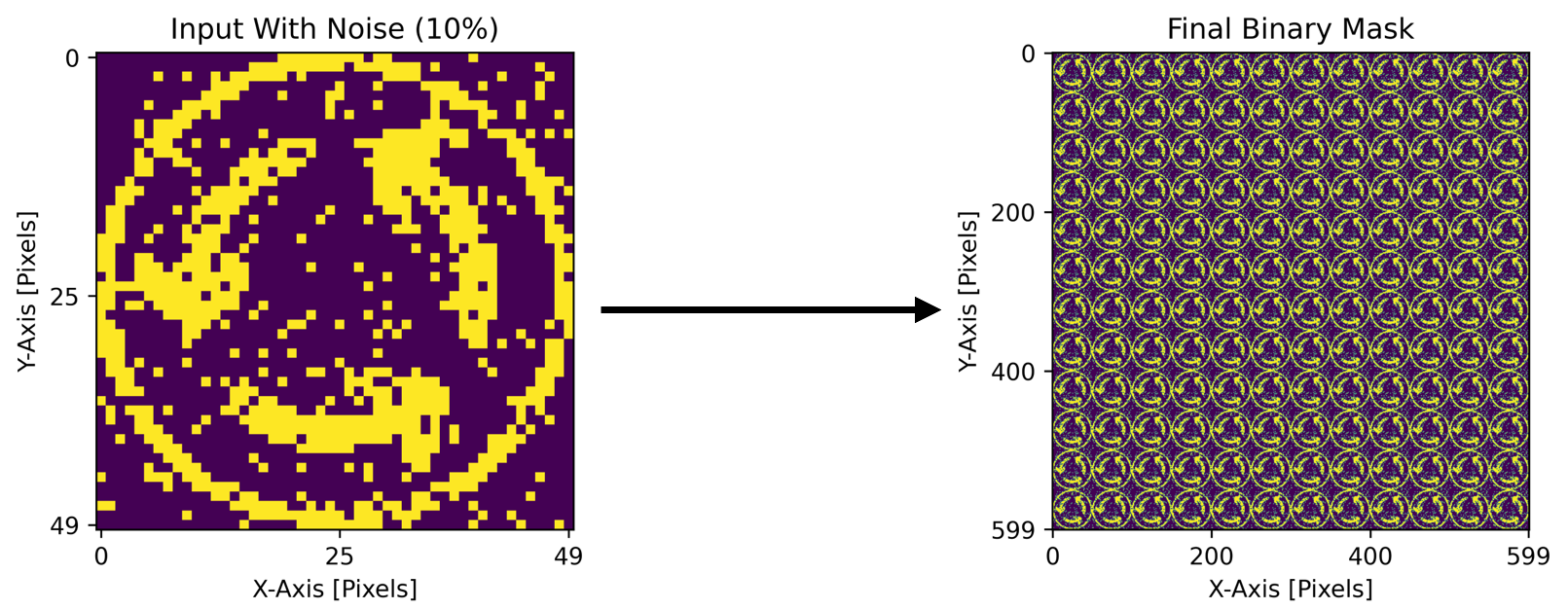}
\end{subfigure}

\caption{\textbf{A,} A random noise that modifies the image in either Constructive [0 → 1] or Destructive [1 → 0] was applied. The percentage (or the amounts of pixels) varies for each test from 1\% to 20\%  with 1\% steps. The Constructive and Destructive noise was shown with blue and red pixels respectively. \textbf{B,} The input with noise is grid wise multiplied with the same grid size as the trained phase mask. The result from element wise multiplication of the final binary mask and the optimized phase mask is then inserted to the SLM to test each noise degree on each image.}
\label{suplFig1}
\end{figure}

\end{document}